\documentclass[10pt]{article}
\usepackage[font={small}, labelsep=period, indention=0.5in]{caption}
\usepackage[OE]{express}
\usepackage{subfig}
\usepackage{float}

\usepackage{graphicx}
\usepackage{lipsum} % just for the example
\usepackage{booktabs}
\usepackage{tabularx}
\usepackage{array}
\newcolumntype{L}[1]{>{\raggedright\let\newline\\\arraybackslash\hspace{0pt}}m{#1}}
\newcolumntype{C}[1]{>{\centering\let\newline\\\arraybackslash\hspace{0pt}}m{#1}}
\newcolumntype{R}[1]{>{\raggedleft\let\newline\\\arraybackslash\hspace{0pt}}m{#1}}

\begin{document}

\title{Efficient ultra-broadband low-resolution astrophotonic spectrographs}

\author{Pradip Gatkine,\authormark{1,2,*} 
Greg Sercel\authormark{2}, 
Nemanja Jovanovic,\authormark{2}, 
Ronald Broeke\authormark{3}, 
Katarzyna Ławniczuk\authormark{3}, 
Marco Passoni\authormark{3},  
Ashok Balakrishnan\authormark{4}, 
Serge Bidnyk\authormark{4}, 
Jielong Yin\authormark{4},
Jeffrey Jewell\authormark{5}, 
J. Kent Wallace\authormark{5}, 
Dimitri Mawet,\authormark{2,5}}

%Nems ORCID [0000-0001-5213-6207]

\address{\authormark{1} Department of Physics \& Astronomy, University of California, Los Angeles (UCLA), 475 Portola Plaza, Los Angeles 90095, USA\\
\authormark{2}Department of Astronomy, California Institute of Technology, 1200 E. California Blvd., Pasadena, CA 91125, USA\\
\authormark{3}Bright Photonics BV, Horsten 1, 5612 AX Eindhoven, The Netherlands\\
\authormark{4} Enablence Technologies Inc., 390 March Road, Ottawa, Ontario, Canada K2K 0G7\\
\authormark{5}Jet Propulsion Laboratory, 4800 Oak Grove Drive, Pasadena, CA 91109, USA}

\email{\authormark{*}pgatkine@astro.ucla.edu, NASA Hubble Fellow} %% email address is required
\homepage{https://gatkine.astro.ucla.edu/} %% author's URL, if desired


\begin{abstract} Broadband low-resolution near-infrared spectrographs in a compact form are crucial for ground- and space-based astronomy and other fields of sensing. Astronomical spectroscopy poses stringent requirements including high efficiency, broad band operation ($>$ 300 nm), and in some cases, polarization insensitivity. We present and compare experimental results from the design, fabrication, and characterization of broadband (1200 - 1650 nm) arrayed waveguide grating (AWG) spectrographs built using the two most promising low-loss platforms - Si$_3$N$_4$ (rectangular waveguides) and doped-SiO$_2$ (square waveguides). These AWGs have a resolving power ($\lambda/\Delta\lambda$) of $\sim$200, free spectral range of $\sim$ 200-350 nm, and a small footprint of $\sim$ 50-100 mm$^2$. The peak overall (fiber-chip-fiber) efficiency of the doped-SiO$_2$ AWG was $\sim$ 79\% (1 dB), and it exhibited a negligible polarization-dependent shift compared to the channel spacing. For Si$_3$N$_4$ AWGs, the peak overall efficiency in TE mode was $\sim$ 50\% (3 dB), and the main loss component was found to be fiber-to-chip coupling losses. These broadband AWGs are key to enabling compact integrations such as multi-object spectrographs or dispersion back-ends for other astrophotonic devices such as photonic lanterns or nulling interferometers. 
% 1. Importance of broadband, low-res spectrographs in Astronomy\\
% 2. Why on-chip? (compactness, compatibility with nulling interferometers, backend for PLs)\\
% 3. We designed and fabricated low-res AWGs to test the performance at a variety of specs that may be needed in astronomy.
% 3. Further compared SiN and doped-SiO2 platforms.\\
% 4. SiN and doped-SiO2: key loss terms (insertion loss, coupling loss, propagation loss)\\
% 4. Doped-SiO2 more suitable for astronomy: larger chips, but lower loss. 
% 5. Comparing their polarization dependence. 
\end{abstract}

 % REPLACE WITH CORRECT OCIS CODES FOR YOUR ARTICLE, MINIMUM OF TWO; Avoid using the OCIS codes for “General” or “General science” whenever possible.
%For a complete list of OCIS codes, visit: https://www.osapublishing.org/oe/submit/ocis/

%%%%%%%%%%%%%%%%%%%%%%% References %%%%%%%%%%%%%%%%%%%%%%%%%
%% This section is commented out and pasted all the way after end document. Please reinstate it here in the final version using bibitem entries. 

%%%%%%%%%%%%%%%%%%%%%%%%%%  body  %%%%%%%%%%%%%%%%%%%%%%%%%%
\section{Introduction \label{sec:intro}}

%\textbf{Scientific motivation for broadband low-res spectrographs:}\\

Low-resolution spectroscopy is an essential tool in astronomy for a myriad of science cases, particularly where the sources are faint, or rapid spectroscopy is desired. Some of these cases include the characterization of exoplanet atmospheres \cite{burrows2014spectra, swain2010ground}, planetary missions \cite{demeo2009extension, kohout2020miniaturized, lantz2020planetary}, and transients such as Gamma-ray bursts, kilonovae, and supernovae \cite{shahbandeh2022carnegie}.
In addition, low-resolution spectroscopy ($\lambda/\delta\lambda \sim 100$, where $\lambda$ = wavelength, $\delta\lambda$ = spectral resolution element) is typically desirable in massively multiplexed (multi-object or integral-field) spectrographs to accommodate a large number of spectra on a limited detector area, particularly for survey science and high-spatial-resolution spectroscopy. Astronomical applications pose a set of challenging requirements including high throughput, broad operational band, and polarization insensitivity. Further, space-based telescopes require the spectrographs to be highly compact, and multi-object spectroscopy requires the spectrographs to be highly replicable. 
% Show the particular importance of low-res broadband spectrographs (not just Astrophotonic spectrographs). Mention HWO.  
%The science cases they address include stellar clusters, active galactic nuclei, galaxy redshift surveys, and cosmology (e.g.: the Dark Energy Spectroscopic Survey).  
%There are diverse applications beyond astronomy, such as Earth-observation and biomedical diagnostics.

On-chip photonic spectrographs are well-suited to address many of these challenges. 
Photonic spectrographs operate in the single-mode regime and thus, are diffraction-limited instruments. Further, these on-chip spectrographs are compact and replicable.
%The resolving power of the slit-based bulk optics spectrographs has a dependence on the angular size of the slit or the source. A larger slit size or source size results in a lower spectral resolution. On the other hand, the resolving power of a given photonic spectrograph is fixed due to the fixed size of the waveguides on the chip, thus ensuring a stable spectral resolution for all observations. 
Unlike conventional bulk optic spectrographs, photonic spectrographs allow design flexibility in terms of routing of light,  placement of spectral channels, and filtering of specific lines (e.g.: atmospheric OH-emission lines \cite{zhu2016arbitrary}), thus catering to a diverse set of science cases. 
Over the last few years, several novel astrophotonic functionalities have been proposed and experimentally demonstrated \cite{jovanovic20232023}, including photonic-lantern-based wavefront sensing \cite{norris2020all}, injection of post-coronagraph light into a single-mode fiber \cite{delorme2021keck, echeverri2023vortex}, 
photonic interferometry \cite{jocou2010gravity}, photonic nulling for high-contrast imaging \cite{martinache2018kernel, norris2020first}, 
photonic spectrographs \cite{cvetojevic2012developing, cvetojevic2012first, gatkine2017arrayed, stoll2020performance, stoll2021design}, Fiber/waveguide-based Bragg gratings for atmospheric OH-suppression\cite{ellis2020first, hu2020integrated},
 spectro-astrometry \cite{kim2022spectroastrometry}, dynamic stabilization of calibration sources \cite{jovanovic2022flattening} and more. An extensive discussion of the challenges in astrophotonics and potential solutions is presented in the Astrophotonics Roadmap \cite{jovanovic20232023}.   These implementations tend to have low-order wavelength dependence. 
Low-resolution astrophotonic spectrographs are highly desirable in significantly enhancing the utility of these implementations by providing on-chip wavelength dispersion. Their diverse applications go well beyond astronomy, such as Earth-observation \cite{van2016high, platt2021ideal} and biomedical diagnostics \cite{seyringer2019compact}.

Arrayed waveguide gratings (AWGs) provide a promising architecture for building astrophotonic spectrographs. 
In the past, low-resolution AWG spectrographs have been demonstrated \cite{gatkine2017} but did not satisfy all three requirements (polarization, bandwidth, throughput) that are essential for astronomy. In this paper, we experimentally investigated three broadband low-resolution  AWGs using commercial 
SiN and SiO$_2$ 
platforms to achieve high-performance, low-resolution spectroscopy for astronomy. We briefly describe the AWG designs in section 2, discuss the measured performance of the fabricated AWGs in section 3, compare these results in section 4, and lay down future directions in section 5. 

% - Exoplanet atmospheres, Earth atmosphere, Planetary missions, \\
% - Cosmology, high-redshift galaxies, spectroscopic surveys at high spatial resolution, 
% - Spectro-interferometers, \\
% - Challenges with the current low-res spectrographs.\\
% Nem - Yes, but limit this to one paragraph. You don't need to convince people we need spectrographs in astronomy or that we need low res. Just why you chose to start in low res regime. 

% For exoplanets, spectro-interferometry, PL-spectroscopy implementations have been proposed. Astrophotonic spectrographs would be of high relevance here. 

% \noindent
% \textbf{Advent of photonics:}\\
% - New measurements enabled by photonics, particularly at high spatial resolution or high-contrast\\
% - Nulling interferometers on a chip: nulling is wavelength-dependent, backend spectrographs needed\\
% - PL-backend spectrographs for spectro-astrometry, stellar nulling\\
% % Nem- list these and cite but don't go into a lot of detail. Keep the message simple and focused on the main points - the low res high performance spectrographs. 

% \noindent
% \textbf{AWGs}: 
% - AWGs is the most promising photonic architecture for spectroscopy.\\
% - State-of-the-art of current broadband AWGs with respect to specific needs in Astronomy (low-loss, polarization insensitive, broadband)\\
% - Current material platforms (commercial) that can help build high-throughput broadband spectrographs (SiN: low-contrast; doped-SiO2)\\

\section{AWG designs and material platforms}

Silicon nitride (Si$_3$N$_4$ core, SiO$_2$ cladding) and Ge-doped-silica (doped SiO$_2$ core, SiO$_2$ cladding) have been shown as two promising material platforms for producing low-loss photonic devices in the near-IR \cite{blumenthal2018silicon, stoll2021design}. Therefore, we explored these material platforms for constructing the low-resolution AWGs. The key design requirements of the AWGs are described below.

The AWGs are required to operate over a broad band
for astronomical applications. 
We chose to focus on a band spanning 1.2 to 1.65 $\mu m$ since it partially covers the astronomical J- (1170-1330 nm) and H-bands (1490-1780 nm) \cite{simons2002mauna}. This waveband has also been the focus of several astrophotonic technologies thanks to the legacy of the telecommunication industry \cite{chen2018emergence}, which makes it the most mature waveband in integrated photonics and, thus, ideal for examining the performance limits. The choice of waveband constrains the waveguide geometry to ensure single-mode operation across the entire band. The spectral resolution required was R $\sim 200$ to ensure both high signal-to-noise-ratio and the wavelength dispersion needed for the science cases described in section \ref{sec:intro}. The free spectral range (FSR) of the AWGs is designed to be $\sim200$ nm to partially cover the span of the astronomical H-band. We further explored an additional design with an FSR of 350 nm for the SiN material to understand potential challenges in scaling these designs to a larger FSR.

In addition, the polarization-dependent shift in wavelength is required to be less than half of the spectral channel separation to ensure that the resolving power degradation for unpolarized light is less than a factor of 2\cite{gatkine2018towards}. However, the polarization dependence requirement can be relaxed by employing off-chip or on-chip broadband polarization splitters and rotators in the future \cite{xu2016ultracompact}.\\  

%\noindent \textbf{Polarization dependence: } This is a soft requirement. \\

\noindent \textbf{Waveguide geometry for SiN:} For polarization dependence, a near-square SiN waveguide could be used, for instance, as offered by Ligentec (a slight trapezoid measuring 800 nm $\times$ 800 nm). However, this waveguide geometry has high-index contrast ($\sim 19\%$). The advantage of high-index contrast is that the modes are highly confined, thus allowing sharper bends (with $R_{bend}~<$ 100 $\mu m$), and thereby, ultra-compact footprint \cite{gatkine2021chip}. The disadvantage of high index contrast is the high differential between the effective index of the waveguide mode and the effective index of the fiber mode, resulting in high fiber-waveguide coupling losses, even with width tapers ($\sim$2.5 dB/facet, \cite{gatkine2021potential}). Such losses are prohibitive in astronomy, and hence, further work is needed to optimize the fiber-waveguide tapers for near-square SiN waveguides. 

On the other hand, ultra-thin waveguides (height $\sim$ 50 nm) are suitable for low propagation and coupling losses. However, they lead to large polarization-dependent losses and polarization-dependent wavelength shifts in the AWG spectral channels due to the weak confinement factor of the TM mode and large footprint (several cm$^2$) due to the large radius of curvature needed to minimize bend losses \cite{dai2011low}. 

Given these constraints, we chose a rectangular waveguide geometry (1000 $\times$ 200 nm) for the SiN AWGs in this paper to minimize the losses at the expense of polarization dependence. The SiN AWGs (\#1 and \#2)  were designed for TE mode. Note that some astronomical applications do require polarization-sensitive spectrographs. For instance, the spectro-interferometers (spectrometers to disperse interference fringes) are typically polarization sensitive since the interference fringes are polarization sensitive.\\

\noindent \textbf{Waveguide geometry for doped-SiO$_2$:} Given the low index contrast of commercially available doped-SiO$_2$ platform ($n_{core}$ = 1.47 and  $n_{cladding}$ = 1.44  gives an index contrast $\Delta~ = ~(n_{core}^{2} - n_{clad}^2)/2n_{core}^{2}$ = $2\%$), square waveguides do not pose the challenge of a large index step between the waveguide and fiber modes, thus allowing a high fiber-to-waveguide coupling efficiency. Therefore, we use a waveguide geometry of 3.4 $\times$ 3.4 $\mu m$, to ensure polarization symmetry.\\ 

% Add single-mode cut-off wavelength if possible. S
%Solver: https://www.siio.eu/eims.html

%\noindent High-contrast SiN

%Talk about waveguide cross-section, tapers,
\noindent
\textbf{Star couplers:} The slab structure between the input waveguides and the arrayed waveguide as well as between the output waveguide and arrayed waveguides is called the star coupler or free propagation region. The shapes of the key sections of the star couplers (i.e., the interface between the slab and the waveguides) are the same in all the AWGs, which are already established \cite{okamoto2010fundamentals}.  The length of the star coupler is the relevant parameter for an AWG. The width of the star coupler only needs to be sufficient to accommodate all the waveguides at the interface of the slab and the arrayed waveguides. AWG \#1 and AWG \#3 have similar specifications for FSR and resolving power. Hence, the lengths of their respective star couplers are similar (1.2 mm and 1.5 mm for AWG \#1 and AWG \#3, respectively) and are simply inversely proportional to their respective effective indices. AWG \#2 has a higher number of output waveguides since the total number of spectral channels is higher, and thus the star-coupler length scales up accordingly (3.15 mm). The AWG design procedure that we followed to arrive at the star coupler and the AWG geometric parameters is well-established \cite{smit1996phasar, okamoto2010fundamentals}.\\

\begin{figure}[h!]
\centering\includegraphics[width=0.75\textwidth]{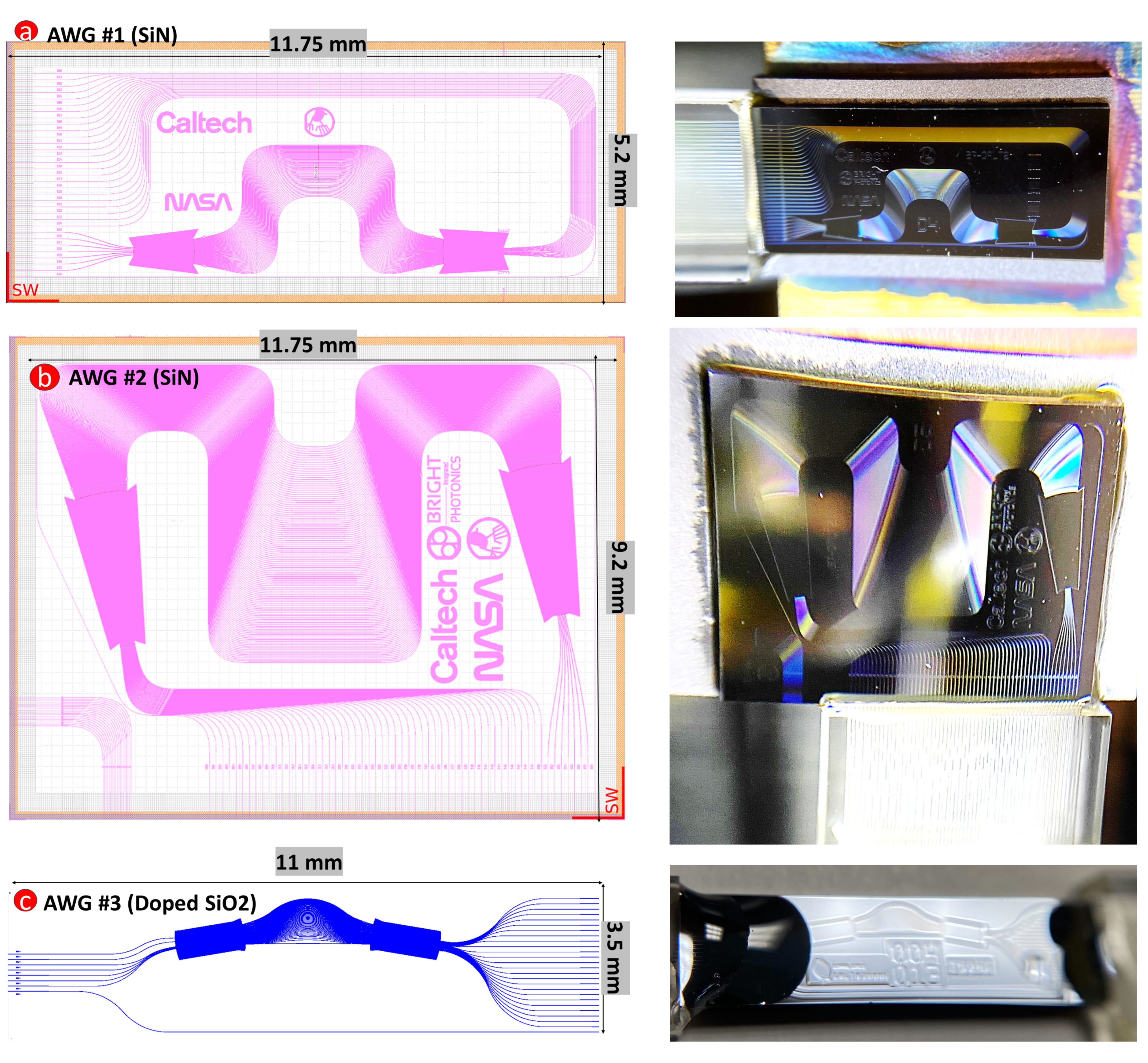}
\caption{The CAD profiles and images of each AWG. AWG \#1 is the top row (11.75 mm $\times$ 5.2 mm), AWG \#2 is the middle row (11.75 mm $\times$ 9.2 mm), and AWG \#3 is at the bottom (11 mm $\times$ 3.5 mm). 
}\label{fig:AWG_CADs}
\end{figure}

\noindent \textbf{Fabrication: } With the waveguide geometries and target resolving power as described above, the properties of the as-designed AWGs are summarized in Table \ref{tab:awg_summary}. Mask designs of the AWGs are shown in Fig. \ref{fig:AWG_CADs}. 
The AWGs presented here have a large FSR, and hence, require a small differential length (= grating order $\times$ $\lambda$) between the arrayed waveguides. Such small differential lengths can be 
accommodated in multiple ways. For low-index-contrast platforms (such as doped-SiO$_2$) with large bend radii, the banana shape (Fig. \ref{fig:AWG_CADs}-bottom) is the most suitable solution for the AWG geometry. For high-index-contrast platforms (such as SiN), where small bending radii are feasible, a folded architecture can be implemented. A folded architecture allows arbitrarily small differential lengths between a large number of arrayed waveguides in a compact form factor. This architecture is used for AWGs \#1 and \#2 (Fig. \ref{fig:AWG_CADs}: top and middle). 
The four bends in the arrayed waveguides in AWG \#1 are used to minimize the discrepancy in the bend loss between the innermost and outermost waveguides. The configuration in AWG \#1 creates a uniform bend loss across all waveguides. This was not needed in AWG \#3 since a larger radius of curvature is used.  The 6 bends in the arrayed waveguides in AWG \#2 are used to accommodate the full AWG delay lines within the footprint of the device. Without those bends, the device footprint will be significantly larger. 

The footprints of AWGs \#1 (SiN FSR = 180 nm), \#2 (SiN FSR = 350 nm), and \#3 (doped-SiO$_2$ FSR = 200 nm) are 11.75 $\times$ 5.2 mm$^2$, 11.75 $\times$ 9.2 mm$^2$, and 11 $\times$ 3.5 mm$^2$.
The SiN AWG was fabricated by Lionix International. The base SiO$_2$ cladding is grown thermally on the silicon substrate followed by low-pressure chemical vapor deposition (LPCVD) deposition of SiN layer (200 nm).  The AWG mask is imprinted using UV contact lithography, followed by dry etching and photoresist removal. Finally, a top layer of  SiO$_2$  cladding is deposited using plasma-enhanced chemical vapor deposition (PECVD). 
The doped-SiO$_2$ AWG was fabricated by Enablence. A Ge-doped silica (core) layer of 3.4 $\mu m$ thickness was deposited on top of a
15 $\mu m$ thick thermally grown SiO$_2$ layer. UV stepper lithography was used to imprint the AWG pattern, followed by dry etching and resist removal. Finally, a SiO$_2$ top cladding layer of 20 $\mu m$ thickness was deposited.

\begin{table}[ht]
\caption{Summary of the characteristics of the broadband low-resolution AWGs.} 
\label{tab:awg_summary}
\begin{center}       
\begin{tabular}{|C{2.5cm}|C{2.5cm}|C{2.5cm}|C{2.5cm}|} 
\hline
\rule[-1ex]{0pt}{3.5ex}   &  AWG\#1 & AWG\#2 & AWG\#3 \\
\hline
%The thing that looks like0pt is 'zero' pt! and not opt :)
%\multicolumn{4}{|c|}{\textbf{Design}}  \\ \hline \vspace{0.1em}                
\rule[-1ex]{0pt}{3.5ex} Material platform  & Si$_3$N$_4$ & Si$_3$N$_4$ & Doped-SiO$_2$\\
\hline
\rule[-1ex]{0pt}{3.5ex}  Waveguide Geometry  & 1000x200 nm & 1000x200 nm & 3400x3400 nm \\
\hline
\rule[-1ex]{0pt}{3.5ex}  Effective index  & TE: 1.58, TM: 1.5  & TE: 1.58, TM: 1.5  & TE and TM: 1.49\\
\hline
\rule[-1ex]{0pt}{3.5ex} Min. R$_\mathrm{curve}$  & 500 $\mu m$ & 500 $\mu m$ & 1500 $\mu m$ \\
\hline
\rule[-1ex]{0pt}{3.5ex} Channel Spacing ($\Delta\lambda$)  & 7.5 nm & 8 nm &  8.75 nm \\
\hline
\rule[-1ex]{0pt}{3.5ex}  Central wavelength  & 1550 nm & 1550 nm & 1550 nm  \\
\hline
\rule[-1ex]{0pt}{3.5ex} Resolving power ($\lambda/\Delta\lambda$)  & 200 & 190 & 175 \\
\hline
\rule[-1ex]{0pt}{3.5ex} FSR  & 180 nm  & 350 nm & 200 nm \\
\hline
\rule[-1ex]{0pt}{3.5ex} Footprint & 11.75$\times$5.2 mm$^2$ & 11.75$\times$9.2 mm$^2$ & 11$\times$3.5 mm$^2$ \\

%\multicolumn{4}{|c|}{\textbf{Characterization}}  \\ \hline \vspace{0.1em}               
%\rule[-1ex]{0pt}{3.5ex}  Fiber used & UHNA3 & UHNA3 & PM1550-XP\\
\hline
%\rule[-1ex]{0pt}{3.5ex}  Setup & Broadband source, Polarization controller, Fiber rotator, AWG, OSA & Broadband source, Polarization controller, Fiber rotator, AWG, OSA & Broadband source, Fiber rotator, AWG, OSA\\
%\hline
\end{tabular}
\end{center}
\end{table}

\section{Results and Discussion}

\subsection{Characterization setup}
The input and output waveguides of AWG\#1 and AWG\#3 were packaged using a polarization-maintaining (PM) fiber array. The input waveguide of the AWG \#2 was packaged with PM fiber and the outputs were packaged with an SMF28 fiber array. The polarization-maintaining fiber of choice was Thorlabs PM1300-XP, which is specified across 1270-1625 nm. 
Note that the tapers used at the output waveguides for AWG \#2 are matched to the SMF28 fiber, and those for AWG \#1 are matched to PM1300-XP. As such, both fiber types have similar mode-field diameters (9.2 microns for SMF28 and 9.3 microns for PM1300XP at 1310 nm), and hence, we do not expect any impact on the comparison between AWG \#1 and AWG \#2. In addition, the optical spectrum analyzer is polarization insensitive, and hence, the polarization state of the AWG output does not impact the measurements.   

The input fiber for the AWG was connected to one of two super luminescent diodes (Thorlabs S5FC1018P for 1200-1400 nm and S5FC1005P for 1400-1650 nm). The output fiber of the AWG was connected to an optical spectrum analyzer (OSA) to analyze the output as a function of wavelength. A fiber-to-fiber response is measured as a reference for source power, which is then used to derive the overall transmission response of the fiber-chip-fiber system.  Note that while AWGs \#1 and \#2 are designed for TE mode only, we measure and report their performance for both TE and TM modes for complete characterization and comparison.

\begin{figure}[h!]
\centering\includegraphics[width=0.99\textwidth]{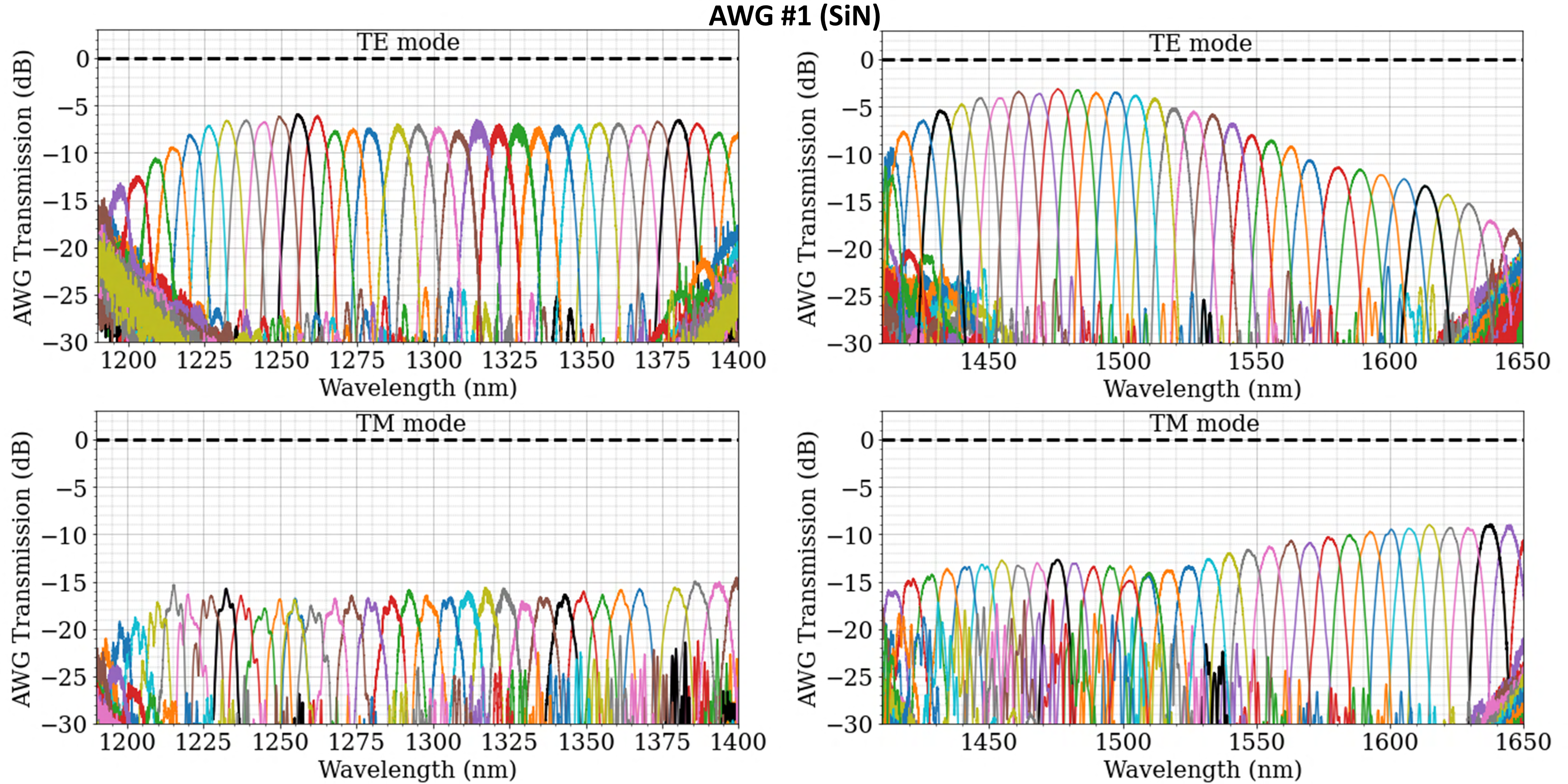}
\caption{
Broadband and polarization-dependent transmission of AWG \#1 (SiN platform). 
\textbf{Top Left:} TE mode transmission response in 1200-1400 nm range.
\textbf{Top Right:} TE mode transmission response in 1400-1600 nm range.
\textbf{Bottom Left:} TM mode transmission response in 1200-1400 nm range.
\textbf{Bottom Right:} TM mode transmission response in 1400-1600 nm range.
}\label{fig:CIF1_D4_Transmission}
\end{figure}

\begin{figure}[h!]
\centering\includegraphics[width=0.99\textwidth]{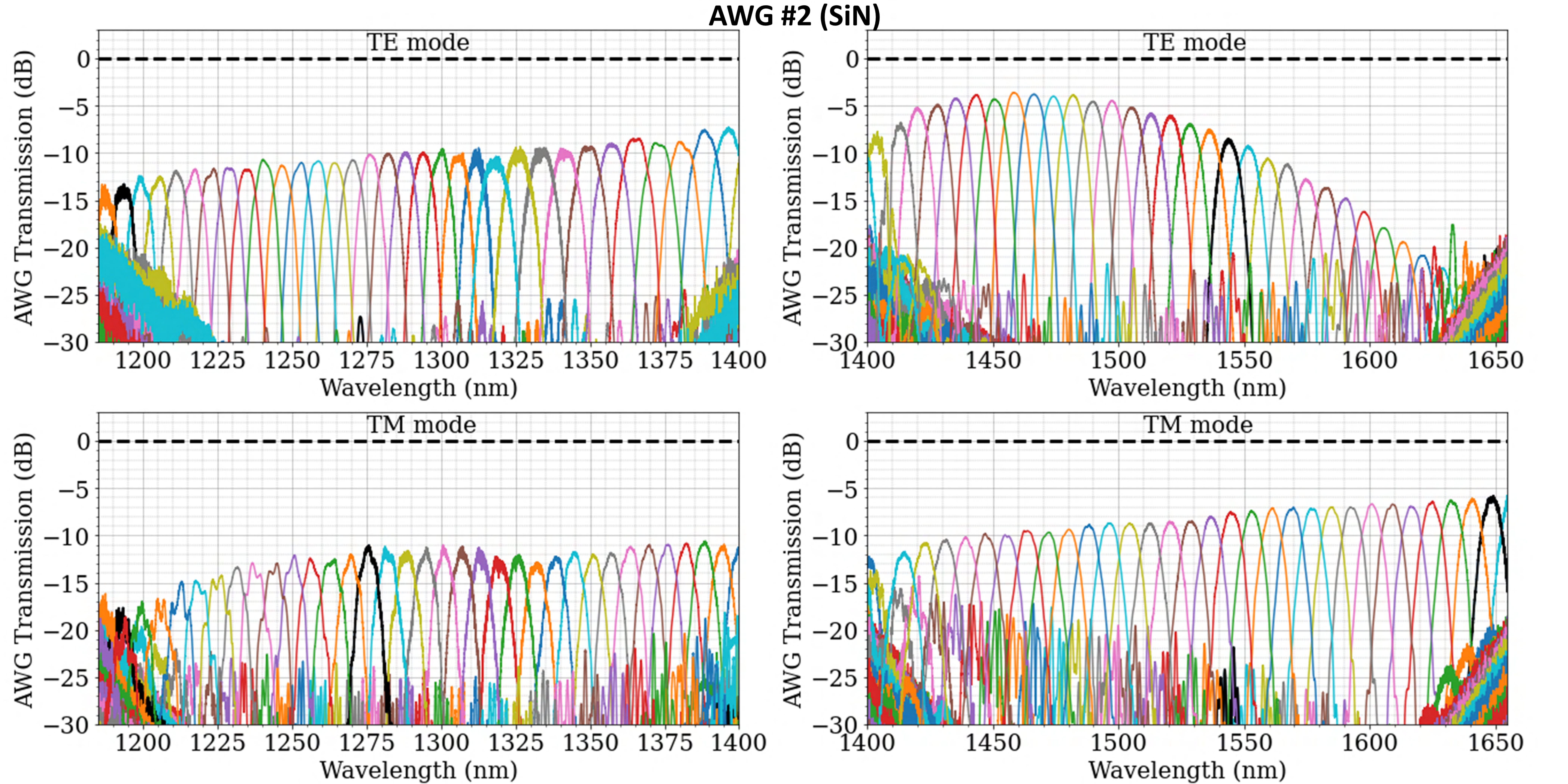}
\caption{Broadband and polarization-dependent transmission of AWG \#2 (SiN platform). 
\textbf{Top Left:} TE mode transmission response in 1200-1400 nm range.
\textbf{Top Right:} TE mode transmission response in 1400-1600 nm range.
\textbf{Bottom Left:} TM mode transmission response in 1200-1400 nm range.
\textbf{Bottom Right:} TM mode transmission response in 1400-1600 nm range.}\label{fig:CIF1_D5_Transmission}
\end{figure}

\begin{figure}[h!]
\centering\includegraphics[width=0.99\textwidth]{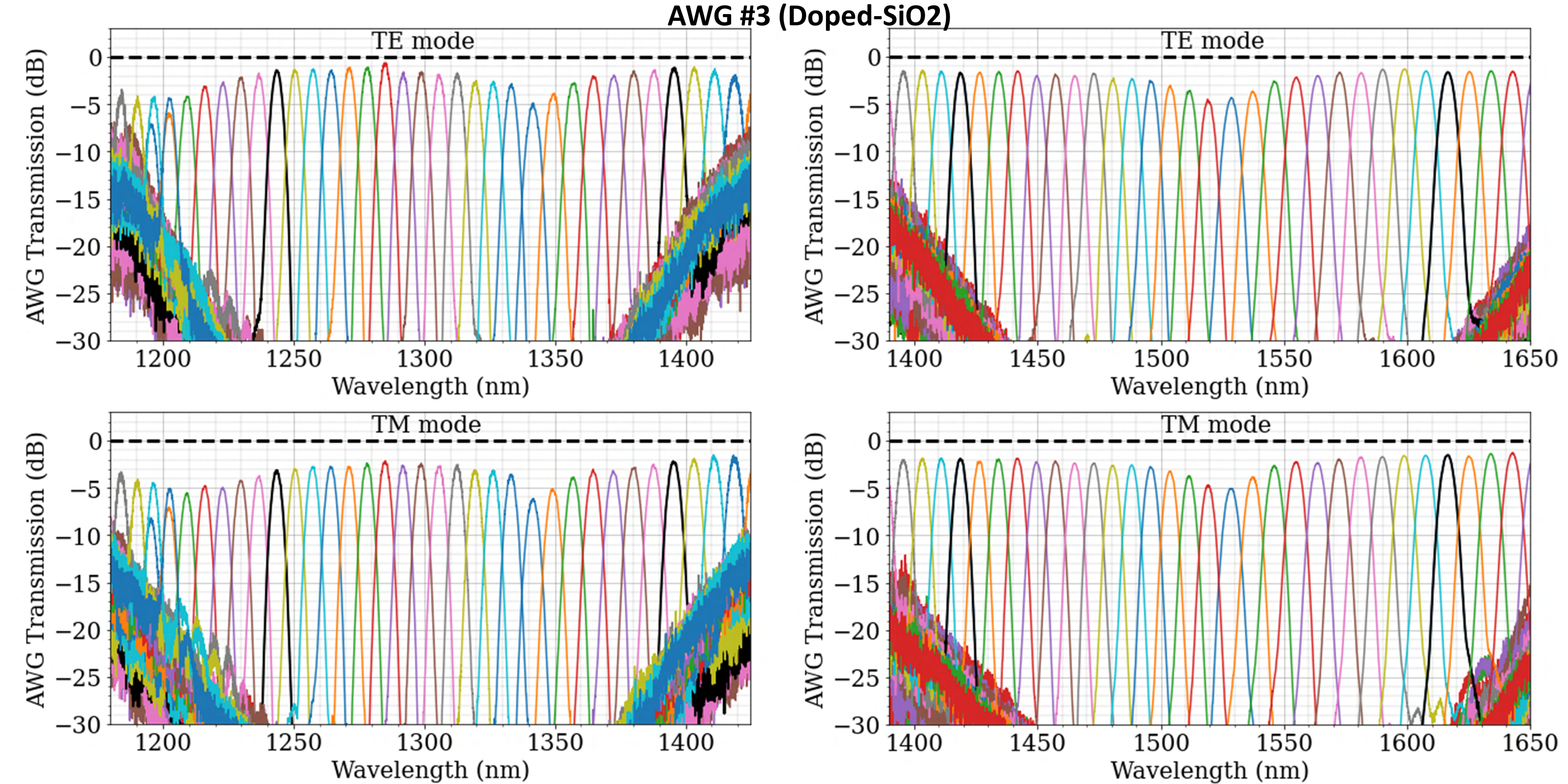}
\caption{\footnotesize Broadband and polarization-dependent transmission of AWG \#3 (SiO2 platform). 
\textbf{Top Left:} TE mode transmission response in 1200-1400 nm range.
\textbf{Top Right:} TE mode transmission response in 1400-1600 nm range.
\textbf{Bottom Left:} TM mode transmission response in 1200-1400 nm range.
\textbf{Bottom Right:} TM mode transmission response in 1400-1600 nm range.
}\label{fig:Enablence_Low_Res}
\end{figure}

\subsection{Transmission response and throughput}\label{sec:Transmission}

The overall transmission of AWGs \#1, \#2, and \# 3 (including fiber-chip coupling) is shown in figures \ref{fig:CIF1_D4_Transmission}, \ref{fig:CIF1_D5_Transmission}, and \ref{fig:Enablence_Low_Res}, respectively. In each figure, the top row shows the TE transmission response and the bottom row shows the TM transmission response across 1200-1400 nm and 1400-1650 nm wavebands. The dotted line at the top indicates 0 dB (100\%) transmission. The spectral channels are shown in different colors. The black trace shows a representative spectral channel for reference and helps visualize the FSR.  The observations are summarized below.\\ 

\begin{figure}[h!]
\centering\includegraphics[width=0.8\textwidth]{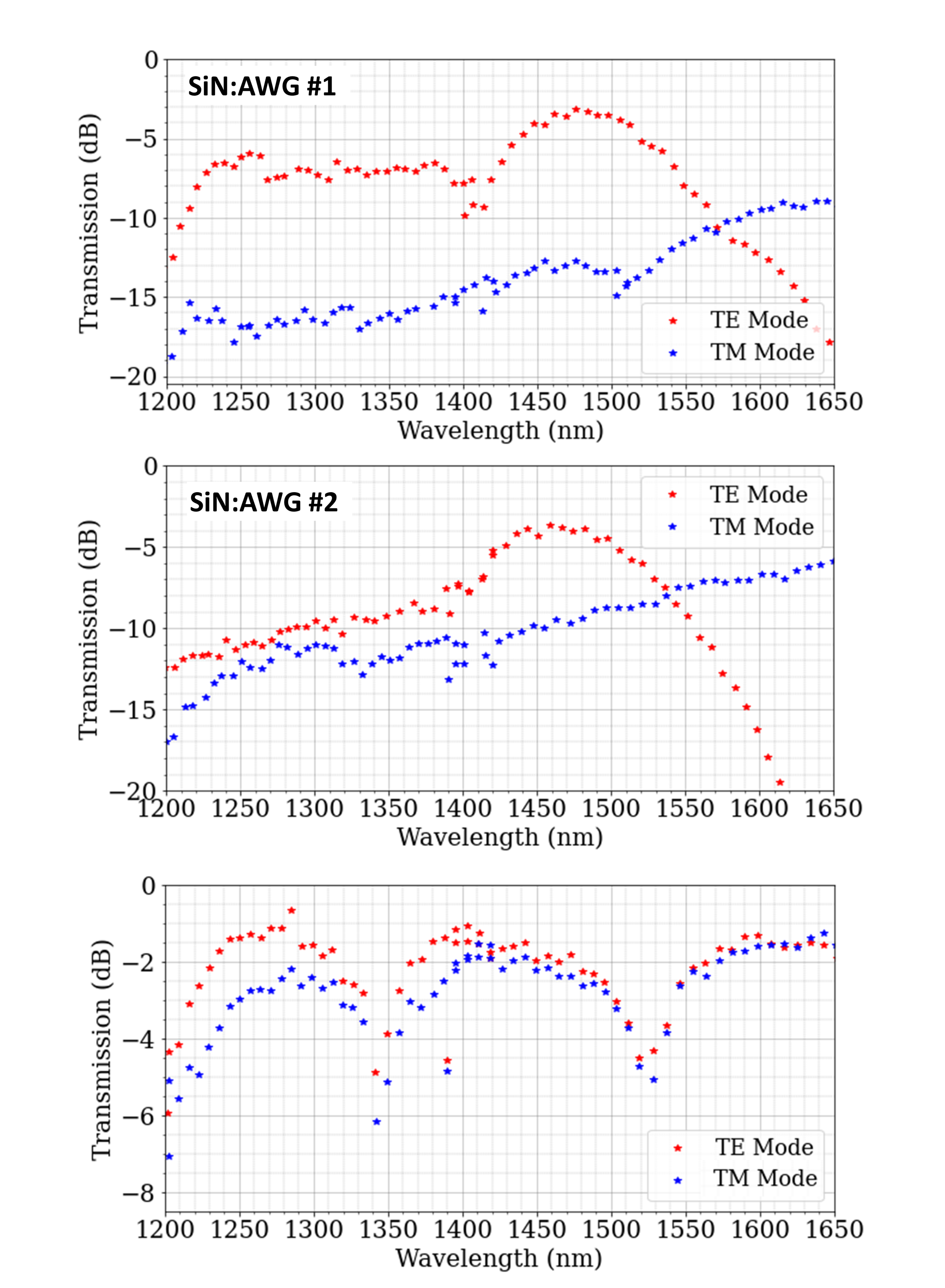}
\caption{\footnotesize TE and TM mode transmission responses across the 1200-1650 nm range for SiN-AWG \#1
(\textbf{top}), SiN-AWG  \#2
(\textbf{middle}), SiO$_2$-AWG  \#3
(\textbf{bottom}).
}\label{fig:Transmission}
\end{figure}

\noindent
\textbf{SiN AWGs:} 
 From the transmission profiles of the spectral channels in figures \ref{fig:CIF1_D4_Transmission} (AWG \#1) and  \ref{fig:CIF1_D5_Transmission} (AWG \#2), it is clear that these AWGs are broadband and operate over a waveband of 1200-1650 nm, albeit with degradation in throughput and/or crosstalk in certain regions depending on the polarization. 
For the TE mode in both AWG \#1 and \#2, the peak transmission of $\sim$ -3 to -3.5 dB occurs around 1450 nm (see Figures \ref{fig:CIF1_D4_Transmission},  \ref{fig:CIF1_D5_Transmission}, and \ref{fig:Transmission}). The transmission rapidly declines at longer wavelengths and flattens out at shorter wavelengths up to 1225 nm (to a level of -7 dB for AWG \#1 and -10 dB for AWG \#2).  On the other hand, for the TM mode in AWGs \#1 and \#2, a roughly monotonic rise is seen in transmission from 1200 nm to 1650 nm.

\begin{figure}[h!]
\centering\includegraphics[width=1.0\textwidth]{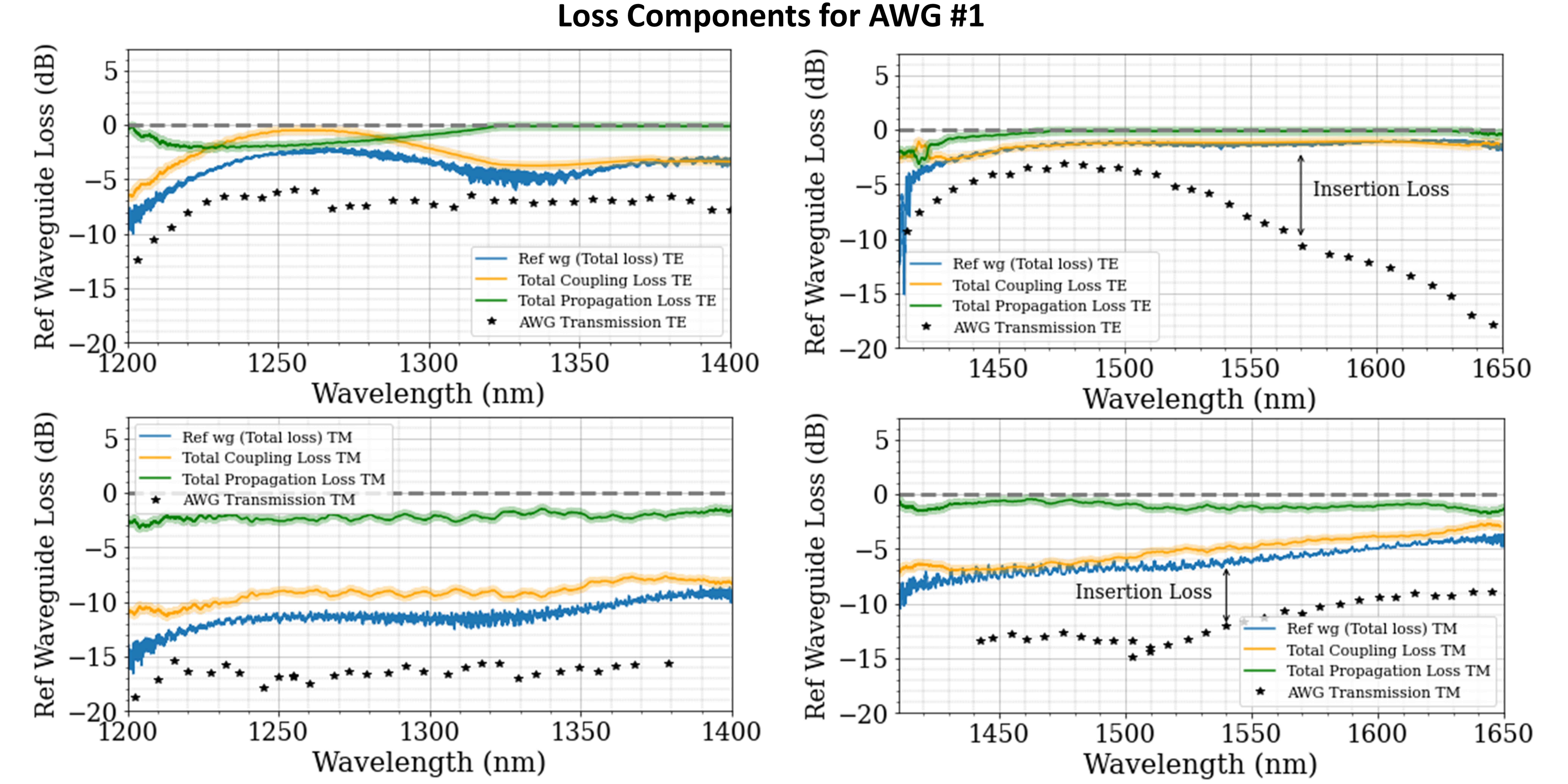}
\caption{\footnotesize TE and TM mode transmission responses across the 1200-1650 nm range for a reference SiN waveguide with length comparable to AWG \#1. The top panel shows TE polarization response and the bottom panel shows TM polarization response. 
}\label{fig:Ref_WG}
\end{figure}

% Add a sentence about Fig 6: Loss decomposition. 
For a qualitative investigation of the loss components, we estimated the wavelength-dependent propagation and coupling losses of AWG \#1. To do that, we measured the fiber-waveguide-fiber transmission for two reference waveguides with different lengths on a chip (from the same wafer and process batch as AWGs \#1 and \#2).  Both the inputs and outputs of these reference waveguides were packaged on the same side of the chip using a V-grove, and hence, we assumed identical coupling loss for both reference waveguides. With this assumption, we used the difference in reference waveguide transmission to estimate the propagation loss as a function of wavelength. By applying this propagation loss to the reference waveguide on AWG \#1 chip, we estimated the coupling loss as a function of wavelength. This decomposition is shown in Fig. \ref{fig:Ref_WG}. In the future, we aim to perform a more precise split of the losses by constructing several reference waveguides  with incremental lengths and measuring their transmission to derive a statistical inference on the loss components (coupling and propagation). 
%We used the reference waveguides of different lengths in the AWG \#1 chip to estimate the wavelength-dependent coupling and propagation components of the loss for AWG \#1 (Fig. \ref{fig:Ref_WG}). 

In Fig. \ref{fig:Ref_WG}, the total reference waveguide loss (fiber-waveguide-fiber) is shown in blue, the total coupling loss (of both facets) is shown in orange, and the total propagation loss is shown in green. The faint bands around the traces indicate the uncertainty.  
The AWG transmission loss is shown with black points. 
The transmission loss for AWG \#1 in the TM (1200 - 1650 nm)  and TE modes (at $\lambda~<$ 1525 nm) is primarily contributed by the coupling loss between fiber-to-waveguide, as shown in Fig. \ref{fig:Ref_WG}. 
Coupling loss shown in Fig. ~\ref{fig:Ref_WG} includes the loss at both input and output facets.   
The TM mode transmission as a function of wavelength for AWG \#2 can also be explained by the coupling loss given the same material platform and waveguide geometry as AWG \#1. Note that there is no anti-reflective (AR) coating on either the fiber or chip facets. AR coating on all the facets will help minimize the Fresnel reflection losses and, thereby, the coupling losses. 

However, the rapid decline in the transmission of AWGs \#1 and \#2 for $\lambda~>$ 1525 nm in the TE mode cannot be explained by the coupling loss since the coupling loss component is only $\sim$ 1 dB. 
In addition, for AWG \#2, in the TE mode, the downward slope at $\lambda~>$ 1525 nm is higher compared to AWG \#1 (as seen in Fig. \ref{fig:Transmission}).  
The only dissimilar elements that could introduce a differential loss between the two AWGs are the propagation losses and the intrinsic insertion loss of the grating (light lost into neighboring spatial orders, not sampled by the output waveguides \cite{smit1996phasar}). Note that the waveguide-to-slab and slab-to-waveguide interfaces are the same in both AWGs, and would thus yield the same loss in both.  
As such, the propagation loss in these SiN waveguides is minimal, as shown in Fig. \ref{fig:Ref_WG}. Thus, it is highly likely that the additional loss at longer wavelengths is due to the intrinsic loss of the grating.

In addition, the transmission of AWG \#2 is $\sim$ 3-4 dB lower than AWG \#1 for $\lambda$ $<$ 1350 nm. We hypothesize a likely reason for this result. The AWG \#2 is twice as large as AWG \#1, and thus the accumulated phase errors due to fabrication errors are proportionally larger. In addition, a given error in the effective waveguide length corresponds to a larger phase error at shorter wavelengths. Thus, it is likely that the shorter wavelengths appear to suffer greater degradation in throughput due to the phase errors. This hypothesis needs to be further investigated through extensive simulations and measurement techniques, such as optical backscatter reflectometry \cite{mechels1999optical}, to pinpoint the source of this loss, but this is beyond the scope of this paper. 

%% Comment: You: Substantiate this with better propagation loss measurements from D3 chip.
%Nemanja Jovanovic: Yeah. How did you conclude this? I think it would not hurt to show one of Marco's plots of the predicted performance for these devices to establish what the design-limited throughput is. Anything beyond that can be attributed to other losses. 

% However, the peak throughput is also quite similar for both AWG1 and AWG2. So, possibly a different, wavelength-sensitive source of error: Bend losses at longer wavelengths and higher scattering loss at longer wavelengths => These would show up as propagation losses. 

%\noindent
%\textbf{Check:} Why this behavior? Is the coupling loss higher at longer wavelengths for TE? Is this a pure artifact of the propagation loss or coupling loss or is it due to the AWG design? (Simulations do not suggest such behavior). Possibly higher bending loss at long wavelengths is the dominant factor for TE and higher scattering loss at short wavelengths is the dominant factor for TM.  

%Crosstalk and higher noise floor in TM.
The typical crosstalk observed for both AWGs \#1 and \#2 in the TE mode is  12-14 dB at $\lambda$ = 1225-1550 nm and 10 dB at $\lambda$ = 1550-1600 nm. For the TM mode, the crosstalk is 12-13 dB at $\lambda$ = 1550-1650 nm and degrades to 5-10 dB at $\lambda$ = 1225-1550 nm due to significant side lobes from non-adjacent channels in the TM mode.\\

%\noindent
%\textbf{Check:} Are SiN waveguides more susceptible to phase errors due to their high index contrast, thus causing higher crosstalk?

\noindent
\textbf{SiO$_2$ AWG:} The transmission response of the SiO$_2$ AWG (AWG \#3) is shown in Figures \ref{fig:Enablence_Low_Res} and \ref{fig:Transmission} (bottom pannel). It is clear that this AWG is broadband and operates over a waveband of 1200-1675 nm without significant degradation in either polarization. 
The peak throughput in the TE mode is $\sim$ 1 dB  (at 1400 and 1275 nm) and 1.5 dB in the TM mode (at 1415 and 1610 nm). 
For both TE and TM modes in AWG \#3, we observe the typical quasi-Gaussian envelopes of the distinct spectral orders (e.g., 1350 to 1540 nm)  that result from the far field electric field distribution of the waveguide geometry at the array-to-output-FPR interface, projected on the AWG's output Rowland circle \cite{smit1996phasar}. 
%Unlike AWGs \#1 and \#2, we observe no monotonic decline in transmission at shorter or longer wavelengths in excess of this envelope.
Unlike AWGs \#1 and \#2, the transmission in AWG \#3 is more uniform across the entire 1200-1650 nm range without significant losses at shorter or longer wavelengths in excess of this envelope.
The non-uniformity loss (difference between central and edge channels in a spectral order envelope) is $\sim$3.5 dB for both TE and TM modes. 

Note that in Fig.~\ref{fig:Enablence_Low_Res}, at $\lambda~<$ 1200 nm, we see spatial wrapping. The next spatial order at the output FPR of the AWG starts to get sampled by the output waveguides, thus we get the constructive interference of the same wavelength at two different output waveguides. Therefore, we define the operational band only down to 1200 nm. This operational waveband can be expanded in the future by increasing the spatial separation between the spatial orders, which can be accomplished by increasing the ratio of the radius of curvature of the free propagation region and the waveguide spacing of the arrayed waveguides \cite{okamoto2010fundamentals}. While expanding the operational waveband, the spectral FSR will need to increase proportionally as we increase the spatial separation between the neighboring spatial orders.

%Crosstalk and higher noise floor in TM.
The crosstalk observed for both  TE and TM modes 25-30 dB across $\lambda$ = 1225-1650 nm with better crosstalk ($\sim$ 30 dB) at $\lambda$ = 1200-1400 nm and slightly worse ($\sim$ 25-27 dB) at $\lambda$ = 1400-1650 nm. This suggests that the phase errors in the SiO$_2$ chip are minimal ($<$ 60$^{\circ}$), thus minimizing the sidelobes, and improving the noise floor contributed by the adjacent and non-adjacent channels \cite{gatkine2021potential}. 

Note that the sharp dips seen in Fig. \ref{fig:Transmission} (particularly bottom panel) are expected in any interference-based spectrograph (photonic or conventional diffraction-grating) since it indicates the boundary of a spectral order. The edge output channels of the spectrograph are off-axis compared to the central output channels, which leads to lower efficiency for edge output channels compared to the central ones. Thus, a dip would appear at the locations of the edge channels for each spectral order. Therefore, the three dips are seen in Fig. \ref{fig:Transmission} (bottom panel) since the 1200-1650 nm range spans roughly 2.5 spectral orders.

\subsection{Resolving power and FSR}\label{sec:Resolution}
\begin{figure}[h!]
\centering\includegraphics[width=0.8\textwidth]{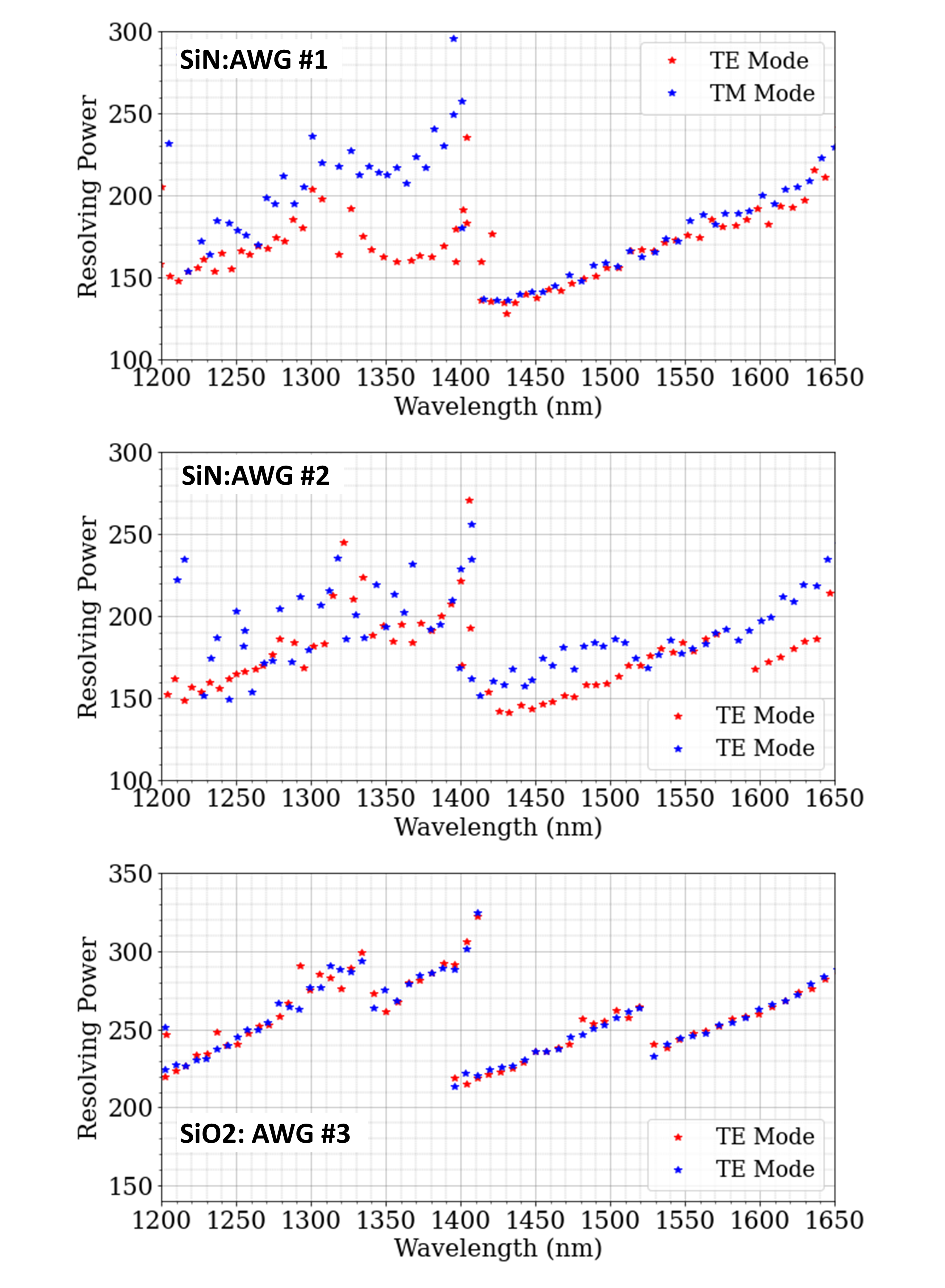}
\caption{\footnotesize TE and TM mode resolving power across the 1200-1650 nm range for SiN-AWG  \#1
(\textbf{top}), SiN-AWG  \#2
(\textbf{middle}), SiO2-AWG  \#3
(\textbf{bottom}).
}\label{fig:Res_power}
\end{figure}

In astronomical spectroscopy, the 3-dB resolving power is defined as $\lambda/\delta\lambda$ where $\delta\lambda$ is the 3-dB width of the spectral channel. We used this definition to estimate the 3-dB resolving powers of each of the fabricated AWGs. It is plotted as a function wavelength in Figure \ref{fig:Res_power}. The typical 3-dB resolving power of AWGs \#1 and \#2 remains in the range of 150-200 across the entire waveband of 1200-1650 nm without a significant difference between TE and TM modes. 
This is consistent with their target resolving powers. 
The scatter in the top and middle panels of Fig. \ref{fig:Res_power} (for SiN AWGs) comes from measurement errors. This is particularly caused in the TM mode since the power in the TM mode is lower at shorter wavelengths due to: a) lower source power at those wavelengths, and b) lower transmission in the TM mode at those wavelengths for the AWG (see Fig. \ref{fig:Transmission}). Therefore, measuring the 3-dB width of the AWG spectral channels incurs a larger error at shorter wavelengths compared to $\lambda$ $>$ 1400 nm.
%($\lambda/\Delta\lambda$ = 190-200), where  $\Delta$ is channel spacing. 
The median 3-dB resolving power of AWG \#3 is between 225 to 275 across the entire operational waveband (1175-1675 nm) with identical performance for both polarizations. Note that the channel spacing of AWG \#3 is consistent with the target of 8.8 nm. The 3-dB width of these channels ($\delta\lambda$) is narrower than the channel spacing due to sharply defined peaks. Therefore, the 3-dB resolving power is higher. 
%This is higher than the as-designed resolving power ($\lambda/\Delta\lambda$) of 175 due to sharply defined peaks.
The sharply defined peaks suggest negligible phase errors (as also deduced in section \ref{sec:Transmission}) \cite{gatkine2021potential} and, thereby, negligible spectral channel broadening (i.e., negligible degradation of spectral resolution). 

The  FSR is given by $\lambda$/(grating spectral order), and hence the wavelength dependence is expected. The FSR for AWG \#1 for the TE (TM) mode is 123 nm (110 nm) at $\lambda$ $\sim$ 1300 nm and 182 nm (162 nm) at $\lambda$ $\sim$ 1550 nm. Note that the TM mode values are mentioned in parentheses. 
For AWG \#2, a single spectral order covers much of the operational waveband, with an FSR of 355 nm in TE mode and 375 nm in TM mode across 1200-1650 nm. For AWG \#3, the TE and TM mode profiles are nearly identical, thus giving an FSR of 150 nm at $\lambda$ $\sim$ 1300 nm and 195 nm at $\lambda$ $\sim$ 1550 nm for both TE and TM modes. All the observed FSR values match the design FSR.

\subsection{Polarization dependence}\label{sec:Polarization}
%We investigated  rectangular- waveguide platform in SiN and a square-waveguide platform in doped-SiO$_2$. Here we report the polarization-dependent shifts in the spectral transmission profiles of the AWGs.\\

\begin{figure}[h!]
\centering\includegraphics[width=0.85\textwidth]{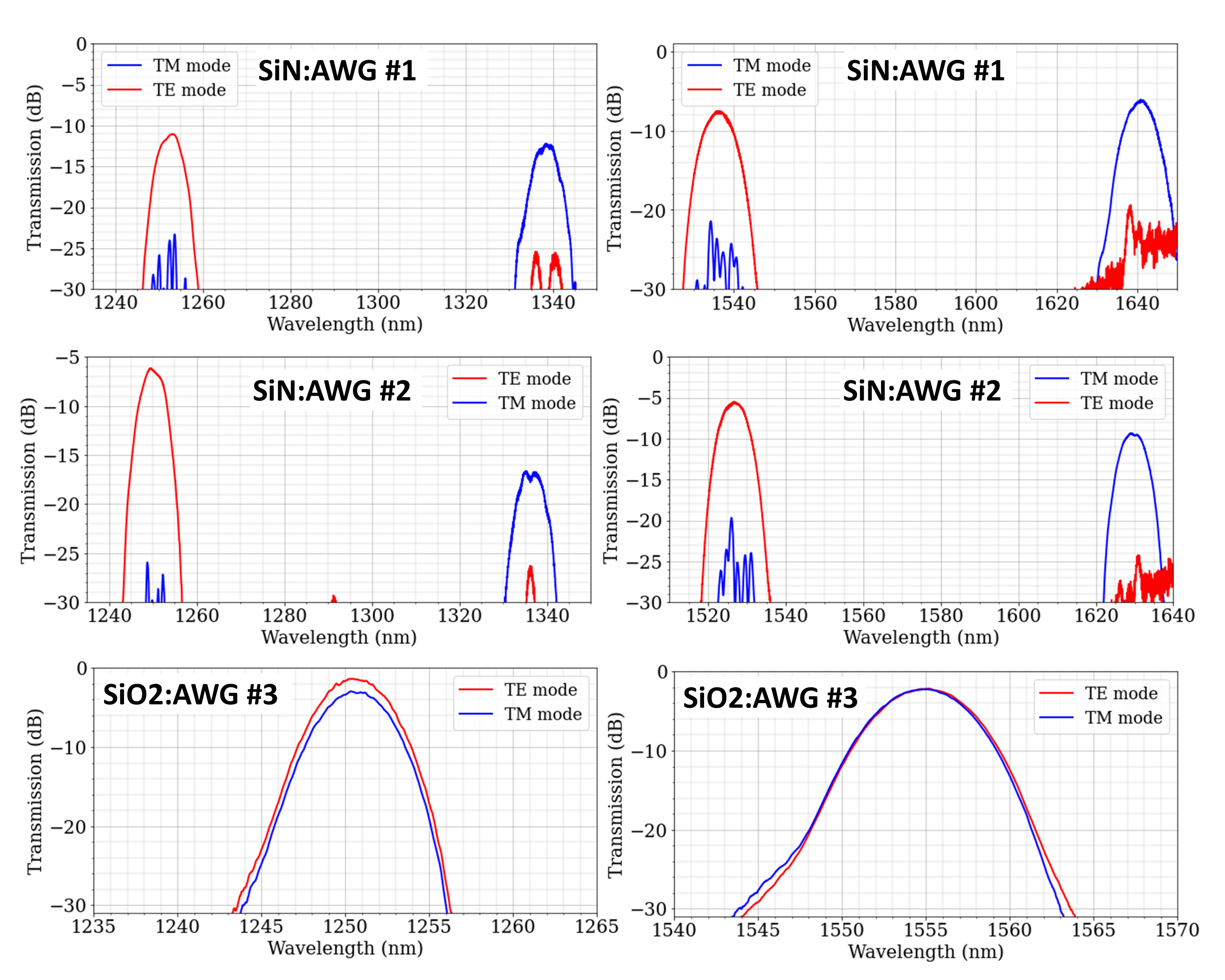}
\caption{\footnotesize Polarization dependence of the 3 chips around 1250 nm and 1550 nm for SiN-AWG  \#1
(\textbf{top row}), SiN-AWG  \#2
(\textbf{middle row}), SiO2-AWG  \#3
(\textbf{bottom row}).}\label{fig:Pol_dependence}
\end{figure}

% The polarization-dependent drift is to be seen as a function of wavelength, and normalized to FSR? 
Note that the SiN AWGs (AWGs \#1 and \#2) were optimized for the TE mode and constructed using rectangular waveguides (1000 $\times$ 200 nm). Therefore, a strong form birefringence is expected in these AWGs. We still study their polarization dependence for the sake of completeness. The measured polarization-dependent wavelength shifts are shown in Fig.~\ref{fig:Pol_dependence}. Both AWG \#1 and \#2 show a polarization-dependent wavelength shift of $\sim 85-90$ nm. This is greater than the expected polarization-dependent shift as calculated from $\Delta n_{eff}~\times~\lambda_{TE}/n_{eff, TE}$ = 78.5 nm at $\lambda$ = 1550 nm\cite{okamoto2010fundamentals}. This is potentially due to the difference between the refractive indices of PECVD-deposited top cladding and thermally grown bottom cladding and needs further investigation beyond the scope of this paper. 

%% Nem comment: I believe this is path length dependent. At the end of the day, the center wavelength for a given polarization will be steered by the OPD accumulated in the array. If TE and TM propagate at different speeds, then the length is the only other variable which will constrain total TE and TM OPD upon propagating through the array. So length definitely matters and the maths here is misleading. 

The SiO$_2$ AWG (AWG \#3) is constructed using square waveguides and, hence, is expected to have negligible form birefringence ($n_{\mathrm{TE}}$ - $n_{\mathrm{TM}}$ $\sim$ 3$\times10^{-5}$). 
Indeed, as seen in Figs. \ref{fig:Enablence_Low_Res} and \ref{fig:Pol_dependence}, the polarization-dependent shift (PD$\lambda$) is $<0.5$ nm, which is negligible compared to the resolution element (i.e., FWHM of 6.22 nm at 1550 nm). Thus, there is no degradation in the resolving power with an unpolarized light source (which is typically the requirement in astronomy). This AWG has a polarization-dependent loss (PDL) of $\sim$1 dB (at $\lambda~<~1400$ nm). A high PD$\lambda$ leads to a degradation in the resolving power by a factor of  1 + (PD$\lambda$/channel spacing).  The PDL, on the other hand, results in degrading the throughput of the AWG. At maximum PDL, one of the polarizations will be lost, leading to only 50\% degradation in throughput (since astronomical sources are typically unpolarized).
Thus, the PD$\lambda$ requirement in astronomy is more stringent compared to PDL and is well-satisfied by the doped-SiO$_2$ AWGs.  
%Indeed, the AWG is observed to be polarization insensitive throughout 1200-1650 nm (see Figs. \ref{fig:Enablence_Low_Res} and \ref{fig:Pol_dependence}). The polarization-dependent shift is $<0.5$ nm, which is negligible compared to the resolution element (i.e., FWHM of 6.22 nm at 1550 nm). Thus, there is no degradation in the resolving power with an unpolarized light source (which is usually the case in astronomy). 

%The dependence of throughput, resolving power, and FSR on polarization is already covered in sections \ref{sec:Transmission} and \ref{sec:Resolution}.

%While the same top-bottom cladding degradation is expected in doped-SiO2 waveguides, the mode is likely less sensitive to this index difference due to the otherwise low-index contrast of the doped-SiO2 waveguides. 

%Provide the theory for expected polarization dependence in the spectral channels due to birefringence. Then infer the birefringence in the respective material platforms using the polarization-dependent shift.   

%- Throughput and resolving power as a function of wavelength and polarization is already covered in other sections. 

%\subsection{High-contrast SiN}

% \paragraph{Resolving power}
% - Both normalized to FSR and simply as a function of wavelength

% \paragraph{Coupling and Propagation Loss}

% \noindent
% - Basic description of SSCs for Ligentec, optimized tapers for SiN-low contrast, and segmented taper for doped-silica (need to measure). A detailed investigation of tapers will be performed in a future paper. In this paper, we measure the coupling and propagation losses of these platforms.\\

% \noindent
% - Coupling losses\\

% \noindent
% How is it measured? Using test waveguides of different lengths. 
% \textbf{Figure} Plot of coupling loss as a function of wavelength for TE and TM polarizations\\

% \noindent
% - Propagation losses\\

% \noindent
% \textbf{Figure} Plot of propagation loss as a function of wavelength for TE and TM polarizations\\

% \paragraph{Overall Throughput}
% Includes coupling and on-chip losses of the AWG.\\ 

% \noindent
% \textbf{Figure} Plot of overall throughput a function of wavelength for TE and TM polarizations\\

%\section{Discussion}
%\subsection{Comparative Summary of Results}

%Main discussion is phase error sensitivity, polarization dependence, and coupling loss dependence on wavelength. The AWG intrinsic loss is higher in SiN (in excess of coupling at shorter and longer wavelengths).\\

%\subsection{Challenges and advantages of broadband astrophotonic spectrographs}

\section{Conclusion and Future Work}
In this paper, we experimentally examined the broadband performance of three low-resolution (R $\sim$ 200) AWGs fabricated using commercial foundries. This exploration is centered around the application for on-chip astronomical spectroscopy. 
AWG \#1 (FSR: 180 nm) and \#2 (FSR: 350 nm) were built using rectangular Si$_3$N$_4$ waveguides, and AWG \#3 (FSR: 200 nm) was constructed using square-shaped doped-SiO$_2$ waveguides. We investigated the transmission response, resolving power, FSR, and polarization dependence of these AWGs as a function of wavelength. We observed that all three AWGs worked over a broad band (1200 $-$ 1650 nm) for both TE and TM polarizations. 

The peak throughput for AWGs \#1 and \#2 was $\sim$ 3.5 dB, and for AWG \#3 was $\sim$ 1 dB for the TE mode. The resolving power obtained was close to the design resolving power for all the AWGs. The PD$\lambda$ was large for AWGs \#1 and \#2 ($\sim$ 80 nm), which was expected due to their rectangular shape and since they were designed for the TE mode.  At the same time, the PD$\lambda$ was found to be negligible for doped-SiO$_2$ waveguides, thanks to their square shape.

Considering the fiber-chip coupling losses, broadband low-loss performance, low crosstalk, and polarization-insensitive performance, we find that doped-SiO$_2$ is an ideally suited platform for low-resolution, broadband on-chip astrophotonic spectroscopy in the astronomical J and H bands (1100-1700 nm). For the SiN platform, the key challenges to be addressed are coupling losses over a broad band and polarization sensitivity. The coupling losses can be alleviated by adding AR coatings to the fiber and chip facets to minimize the Fresnel reflection losses. In addition, optimizing the tapers to lower the effective index at the waveguide facet and using ultra-high numerical aperture fibers can result in better mode matching, and, thus, minimal coupling losses \cite{zhu2016ultrabroadband}.  
For a high-efficiency fiber-waveguide coupling in SiN across 1200$-$1700 nm, efficient spot-size converters, such as those proposed by \cite{yao2020broadband} and \cite{bhandari2020compact} need to be experimentally demonstrated over a broad band.    
The SiN platform offers a smaller radius of curvature thanks to its high index contrast. Hence, a smaller footprint can be achieved by using appropriate folding of the waveguides in the array (as shown for AWG \#2). This is particularly pertinent for AWGs requiring longer delay lines (either due to large FSR or large resolving power). This is the key advantage of SiN devices compared to doped-SiO$_2$ devices. 
The polarization sensitivity can be resolved by using near-square waveguides (such as those offered by Ligentec) or by employing off-chip or on-chip broadband polarization splitters and rotators \cite{xu2016ultracompact} or by tuning the angles of incidence of the respective polarizations to compensate the PD$\lambda$ \cite{han2019polarization} in the future. 
%For a broad band and efficient coupling in SiN across 1200$-$1700 nm, efficient spot-size converters, such as those proposed by \cite{yao2020broadband} and \cite{bhandari2020compact} need to be experimentally demonstrated over a broad band.    

%- Thus, from a packaging perspective and polarization insensitivity SiO2 is a better platform for low-res spectroscopy until the coupling and polarization issues are addressed with SiN (perhaps using square waveguides or other specially engineered waveguides such as the Triplex platform (estimate the TE TM shift for Triplex) or polarizer-rotator on/off-chip). 

However, there are still further developments that need to be undertaken with the doped-SiO$_2$-based AWGs before they are employed for astronomical spectroscopy. Since astronomical sources are extremely faint, collecting light from the entire waveband without any gaps is crucial. Thus, the spectral dropout in between neighboring spectral channels needs to be minimized (see Fig. 2 in \cite{gatkine2021potential} for more details). This can be done by constructing the channel profiles such that the neighboring channels overlap at their 3-dB (50 \%) point. With this, we can eliminate the spectral dropout since the total light sampled by combining the two neighboring channels will be 100 \%. Any residual ripple can be easily calibrated out. Such a construction would yield the same performance as conventional astronomical gratings (eg: Volume-phase holography (VPH) gratings), where the 3-dB point of the point-spread function is used to define the spectral channels. 
While this would increase the crosstalk between the spectral channels, the increased crosstalk is not a significant concern for astronomy, since, in conventional astronomical spectrographs, the 3-dB width is typically used to define the spectral channels, and crosstalk requirements are less stringent.
The current SiO$_2$ AWG was not designed for this specification, and thus, the neighboring channels overlap at $\sim$6-dB point at 1550 nm, leading to a net spectral dropout of  3 dB (= 6 dB - 3 dB). In addition, achieving the spectral dispersion on a flat focal plane, instead of a Rowland circle in conventional AWGs, allows dicing of the AWG along the focal plane. This enables imaging of the entire focal plane on the detector without any discretization or losses due to sampling by output waveguides and is, thus, useful in astronomy. This can be achieved using a three-stigmatic-point AWG design (as previously shown in \cite{hu2020ultra, stoll2021design, zhan2023design}).  
These will be the milestones for a future paper. 

In addition, the non-uniformity loss of the AWG (the difference between the throughputs at the central and edge channels) needs to be minimized. For AWG \#3, it is about 2.5-3 dB. This is comparable to the non-uniformity loss ($\sim$2-4 dB) seen in short-wave infrared (SWIR) VPH gratings, which are typically used in astronomy for high-throughput spectroscopy (see Fig. 4 in \cite{ishikawa2018comprehensive}, 
\cite{ebizuka2017novel}, and \cite{VPH}). 
The non-uniformity loss happens due to the far-field illumination pattern of the waveguide at the interface of the waveguide array and output slab. To minimize the non-uniformity loss, the waveguide geometry at the interface needs to be optimized for a flatter far-field illumination pattern. This can be achieved in the future by introducing nanowire waveguides at the slab-waveguide interface \cite{yuan2022silicon}, among other approaches. Finally, the AWG FSR needs to be extended from the current  200 nm to at least $\sim$ 300 nm to entirely cover the astronomical J-band (1170-1330 nm) or  H-band (1490-1780 nm) without cross-dispersion \cite{simons2002mauna}. This will be the target of a future design, similar to the specifications of AWG \#2, but with doped-SiO$_2$ platform. The footprint of such an AWG is anticipated to be two times the size of AWG \#3. However, even with the increased size, the phase errors will remain $<$ 60$^{\circ}$ (based on the current performance of AWG \#3), and thus, will not degrade the AWG throughput. 
Such low-loss, broadband, low-resolution on-chip astrophotonic spectrographs could prove to be a valuable technology for various capabilities such as spectroscopy of directly imaged exoplanets, spectro-interferometry, spectro-astrometry, and so on for the upcoming space-based Habitable Worlds Observatory. 

%Further challenges to be addressed in SiO2:

%- Broadband polarization insensitivity is needed. \\
%- Doing the J+H band device in SiO2 for HWO. \\
%- Flat focal plane AWG at this resolution for continuous sampling of the output spectrum. \\
%- Low non-uniformity loss is needed in SiO2. What should we do for that? by introducing nanowire waveguides at the slab-waveguide interface in the arrayed \cite{yuan2022silicon} 

%%%%%%%%%%%%%

\section*{Funding}
Support for P Gatkine was provided by NASA through the NASA Hubble Fellowship Grant
HST-HF2-51478.001-A awarded by the Space Telescope Science Institute, which is operated by the
Association of Universities for Research in Astronomy, incorporated, under NASA Contract NAS5-26555.
This work was supported by the Wilf Family Discovery Fund in Space and Planetary Science, funded by the Wilf Family Foundation, as well as the support from Keck Institute for
Space Studies at Caltech. Some of this research was carried out at Caltech and the Jet Propulsion Laboratory
and funded through the President’s and Director’s Research \& Development Fund program. This work was supported by NASA through the Center Innovation Fund.  

\section*{Acknowledgments}
The authors would like to thank the staff at Lionix 
%and Enablence 
for the development of some of the photonic chips presented herein.

\section*{Disclosures}
The authors declare no conflicts of interest.

\section*{Data availability}
Data underlying the results presented in this paper are not publicly available at this time but may be obtained from the authors upon reasonable request.\\

\bibliography{report}

\begin{thebibliography}{10}
\newcommand{\enquote}[1]{``#1''}

\bibitem{burrows2014spectra}
A.~S. Burrows, \enquote{Spectra as windows into exoplanet atmospheres,}
  Proceedings of the National Academy of Sciences \textbf{111}, 12601--12609
  (2014).

\bibitem{swain2010ground}
M.~R. Swain, P.~Deroo, C.~A. Griffith, G.~Tinetti, A.~Thatte, G.~Vasisht,
  P.~Chen, J.~Bouwman, I.~J. Crossfield, D.~Angerhausen \emph{et~al.},
  \enquote{A ground-based near-infrared emission spectrum of the exoplanet hd
  189733b,} Nature \textbf{463}, 637--639 (2010).

\bibitem{demeo2009extension}
F.~E. DeMeo, R.~P. Binzel, S.~M. Slivan, and S.~J. Bus, \enquote{An extension
  of the bus asteroid taxonomy into the near-infrared,} Icarus \textbf{202},
  160--180 (2009).

\bibitem{kohout2020miniaturized}
T.~Kohout and A.~N{\"a}sil{\"a}, \enquote{Miniaturized spectral imaging
  instrumentation for planetary exploration,} Tech. rep., Copernicus Meetings
  (2020).

\bibitem{lantz2020planetary}
C.~Lantz, F.~Poulet, D.~Loizeau, L.~Riu, C.~Pilorget, J.~Carter, H.~Dypvik,
  F.~Rull, and S.~C. Werner, \enquote{Planetary terrestrial analogues library
  project: 1. characterization of samples by near-infrared point spectrometer,}
  Planetary and Space Science \textbf{189}, 104989 (2020).

\bibitem{shahbandeh2022carnegie}
M.~Shahbandeh, E.~Hsiao, C.~Ashall, J.~Teffs, P.~Hoeflich, N.~Morrell,
  M.~Phillips, J.~Anderson, E.~Baron, C.~Burns \emph{et~al.}, \enquote{Carnegie
  supernova project-ii: near-infrared spectroscopy of stripped-envelope
  core-collapse supernovae,} The Astrophysical Journal \textbf{925}, 175
  (2022).

\bibitem{zhu2016arbitrary}
T.~Zhu, Y.~Hu, P.~Gatkine, S.~Veilleux, J.~Bland-Hawthorn, and M.~Dagenais,
  \enquote{Arbitrary on-chip optical filter using complex waveguide {Bragg}
  gratings,} Applied Physics Letters \textbf{108}, 101104 (2016).

\bibitem{jovanovic20232023}
N.~Jovanovic, P.~Gatkine, N.~Anugu, R.~Amezcua-Correa, R.~Basu~Thakur,
  C.~Beichman, C.~Bender, J.-P. Berger, A.~Bigioli, J.~Bland-Hawthorn
  \emph{et~al.}, \enquote{2023 astrophotonics roadmap: pathways to realizing
  multi-functional integrated astrophotonic instruments,} Journal of Physics:
  Photonics  (2023).

\bibitem{norris2020all}
B.~R. Norris, J.~Wei, C.~H. Betters, A.~Wong, and S.~G. Leon-Saval, \enquote{An
  all-photonic focal-plane wavefront sensor,} Nature Communications
  \textbf{11}, 5335 (2020).

\bibitem{delorme2021keck}
J.-R. Delorme, N.~Jovanovic, D.~Echeverri, D.~Mawet, J.~Kent~Wallace, R.~D.
  Bartos, S.~Cetre, P.~Wizinowich, S.~Ragland, S.~Lilley \emph{et~al.},
  \enquote{Keck planet imager and characterizer: a dedicated single-mode fiber
  injection unit for high-resolution exoplanet spectroscopy,} Journal of
  Astronomical Telescopes, Instruments, and Systems \textbf{7}, 035006--035006
  (2021).

\bibitem{echeverri2023vortex}
D.~Echeverri, J.~Xuan, N.~Jovanovic, G.~Ruane, J.-R. Delorme, D.~Mawet,
  B.~Mennesson, E.~Serabyn, J.~K. Wallace, J.~Wang \emph{et~al.},
  \enquote{Vortex fiber nulling for exoplanet observations: implementation and
  first light,} Journal of Astronomical Telescopes, Instruments, and Systems
  \textbf{9}, 035002--035002 (2023).

\bibitem{jocou2010gravity}
L.~Jocou, K.~Perraut, A.~Nolot, J.-P. Berger, T.~Moulin, P.~Labeye, S.~Lacour,
  G.~Perrin, J.-B. Lebouquin, H.~Bartko \emph{et~al.}, \enquote{The gravity
  integrated optics beam combination,} in \enquote{Optical and Infrared
  Interferometry II,} , vol. 7734 (SPIE, 2010), vol. 7734, pp. 1109--1120.

\bibitem{martinache2018kernel}
F.~Martinache and M.~J. Ireland, \enquote{Kernel-nulling for a robust direct
  interferometric detection of extrasolar planets,} Astronomy \& Astrophysics
  \textbf{619}, A87 (2018).

\bibitem{norris2020first}
B.~R. Norris, N.~Cvetojevic, T.~Lagadec, N.~Jovanovic, S.~Gross, A.~Arriola,
  T.~Gretzinger, M.-A. Martinod, O.~Guyon, J.~Lozi \emph{et~al.},
  \enquote{First on-sky demonstration of an integrated-photonic nulling
  interferometer: the glint instrument,} Monthly Notices of the Royal
  Astronomical Society \textbf{491}, 4180--4193 (2020).

\bibitem{cvetojevic2012developing}
N.~Cvetojevic, N.~Jovanovic, J.~Lawrence, M.~Withford, and J.~Bland-Hawthorn,
  \enquote{Developing arrayed waveguide grating spectrographs for multi-object
  astronomical spectroscopy,} Optics express \textbf{20}, 2062--2072 (2012).

\bibitem{cvetojevic2012first}
N.~Cvetojevic, N.~Jovanovic, C.~Betters, J.~Lawrence, S.~Ellis, G.~Robertson,
  and J.~Bland-Hawthorn, \enquote{First starlight spectrum captured using an
  integrated photonic micro-spectrograph,} Astronomy \& Astrophysics
  \textbf{544}, L1 (2012).

\bibitem{gatkine2017arrayed}
P.~Gatkine, S.~Veilleux, Y.~Hu, J.~Bland-Hawthorn, and M.~Dagenais,
  \enquote{Arrayed waveguide grating spectrometers for astronomical
  applications: new results,} Optics Express \textbf{25}, 17918--17935 (2017).

\bibitem{stoll2020performance}
A.~Stoll, K.~Madhav, and M.~Roth, \enquote{Performance limits of astronomical
  arrayed waveguide gratings on a silica platform,} Optics Express \textbf{28},
  39354--39367 (2020).

\bibitem{stoll2021design}
A.~Stoll, K.~Madhav, and M.~Roth, \enquote{Design, simulation and
  characterization of integrated photonic spectrographs for astronomy {II}:
  low-aberration generation-{II} {AWG} devices with three stigmatic points,}
  Optics Express \textbf{29}, 36226--36241 (2021).

\bibitem{ellis2020first}
S.~Ellis, J.~Bland-Hawthorn, J.~Lawrence, A.~Horton, R.~Content, M.~Roth,
  N.~Pai, R.~Zhelem, S.~Case, E.~Hernandez \emph{et~al.}, \enquote{First
  demonstration of oh suppression in a high-efficiency near-infrared
  spectrograph,} Monthly Notices of the Royal Astronomical Society
  \textbf{492}, 2796--2806 (2020).

\bibitem{hu2020integrated}
Y.~Hu, S.~Xie, J.~Zhan, Y.~Zhang, S.~Veilleux, and M.~Dagenais,
  \enquote{Integrated arbitrary filter with spiral gratings: Design and
  characterization,} Journal of Lightwave Technology  (2020).

\bibitem{kim2022spectroastrometry}
Y.~J. Kim, S.~Sallum, J.~Lin, Y.~Xin, B.~Norris, C.~Betters, S.~Leon-Saval,
  J.~Lozi, S.~Vievard, P.~Gatkine \emph{et~al.}, \enquote{Spectroastrometry
  with photonic lanterns,} in \enquote{Ground-based and Airborne
  Instrumentation for Astronomy IX,} , vol. 12184 (SPIE, 2022), vol. 12184, pp.
  1391--1402.

\bibitem{jovanovic2022flattening}
N.~Jovanovic, P.~Gatkine, B.~Shen, M.~Gao, N.~Cvetojevic, K.~{\L}awniczuk,
  R.~Broeke, C.~Beichman, S.~Leifer, J.~Jewell \emph{et~al.},
  \enquote{Flattening laser frequency comb spectra with a high dynamic range,
  broadband spectral shaper on-a-chip,} Optics Express \textbf{30},
  36745--36760 (2022).

\bibitem{van2016high}
L.~van~der Wal, B.~de~Goeij, R.~Jansen, J.~Oosterling, and B.~Snijders,
  \enquote{High-grade, compact spectrometers for earth observation from
  smallsats,} in \enquote{Remote Sensing Technologies and Applications in Urban
  Environments,} , vol. 10008 (SPIE, 2016), vol. 10008, pp. 37--49.

\bibitem{platt2021ideal}
U.~Platt, T.~Wagner, J.~Kuhn, and T.~Leisner, \enquote{The “ideal”
  spectrograph for atmospheric observations,} Atmospheric Measurement
  Techniques \textbf{14}, 6867--6883 (2021).

\bibitem{seyringer2019compact}
D.~Seyringer, M.~Sagmeister, A.~Maese-Novo, M.~Eggeling, E.~Rank, J.~Edlinger,
  P.~Muellner, R.~Hainberger, W.~Drexler, J.~Kraft \emph{et~al.},
  \enquote{Compact and high-resolution 256-channel silicon nitride based
  awg-spectrometer for oct on a chip,} in \enquote{2019 21st International
  Conference on Transparent Optical Networks (ICTON),}  (IEEE, 2019), pp. 1--4.

\bibitem{gatkine2017}
P.~{Gatkine}, S.~{Veilleux}, Y.~{Hu}, J.~{Bland-Hawthorn}, and M.~{Dagenais},
  \enquote{{Arrayed waveguide grating spectrometers for astronomical
  applications: new results},} Optics Express \textbf{25}, 17918 (2017).

\bibitem{blumenthal2018silicon}
D.~J. Blumenthal, R.~Heideman, D.~Geuzebroek, A.~Leinse, and C.~Roeloffzen,
  \enquote{Silicon nitride in silicon photonics,} Proceedings of the IEEE
  \textbf{106}, 2209--2231 (2018).

\bibitem{simons2002mauna}
D.~Simons and A.~Tokunaga, \enquote{The mauna kea observatories near-infrared
  filter set. i. defining optimal 1--5 micron bandpasses,} Publications of the
  Astronomical Society of the Pacific \textbf{114}, 169 (2002).

\bibitem{chen2018emergence}
X.~Chen, M.~M. Milosevic, S.~Stankovi{\'c}, S.~Reynolds, T.~D. Bucio, K.~Li,
  D.~J. Thomson, F.~Gardes, and G.~T. Reed, \enquote{The emergence of silicon
  photonics as a flexible technology platform,} Proceedings of the IEEE
  \textbf{106}, 2101--2116 (2018).

\bibitem{gatkine2018towards}
P.~Gatkine, S.~Veilleux, Y.~Hu, J.~Bland-Hawthorn, and M.~Dagenais,
  \enquote{Towards a multi-input astrophotonic {AWG} spectrograph,} in
  \enquote{Advances in Optical and Mechanical Technologies for Telescopes and
  Instrumentation III,} , vol. 10706 (International Society for Optics and
  Photonics, 2018), vol. 10706, p. 1070656.

\bibitem{xu2016ultracompact}
Y.~Xu and J.~Xiao, \enquote{Ultracompact and high efficient silicon-based
  polarization splitter-rotator using a partially-etched subwavelength grating
  coupler,} Scientific reports \textbf{6}, 27949 (2016).

\bibitem{gatkine2021chip}
P.~Gatkine, N.~Jovanovic, J.~Jewell, J.~K. Wallace, and D.~Mawet, \enquote{An
  on-chip astrophotonic spectrograph with a resolving power of 12,000,} in
  \enquote{UV/Optical/IR Space Telescopes and Instruments: Innovative
  Technologies and Concepts X,} , vol. 11819 (SPIE, 2021), vol. 11819, pp.
  171--180.

\bibitem{gatkine2021potential}
P.~Gatkine, N.~Jovanovic, C.~Hopgood, S.~Ellis, R.~Broeke, K.~{\L}awniczuk,
  J.~Jewell, J.~K. Wallace, and D.~Mawet, \enquote{Potential of commercial
  {SiN} {MPW} platforms for developing mid/high-resolution integrated photonic
  spectrographs for astronomy,} Applied Optics \textbf{60}, D15--D32 (2021).

\bibitem{dai2011low}
D.~Dai, Z.~Wang, J.~F. Bauters, M.-C. Tien, M.~J. Heck, D.~J. Blumenthal, and
  J.~E. Bowers, \enquote{Low-loss silicon nitride arrayed-waveguide grating
  (de) multiplexer using nano-core optical waveguides,} Optics express
  \textbf{19}, 14130--14136 (2011).

\bibitem{okamoto2010fundamentals}
K.~Okamoto, \emph{Fundamentals of optical waveguides} (Academic press, 2010).

\bibitem{smit1996phasar}
M.~K. Smit and C.~Van~Dam, \enquote{Phasar-based wdm-devices: Principles,
  design and applications,} IEEE Journal of Selected Topics in Quantum
  Electronics, 2 (2)  (1996).

\bibitem{mechels1999optical}
S.~Mechels, K.~Takada, and K.~Okamoto, \enquote{Optical low-coherence
  reflectometer for measuring wdm components,} IEEE Photonics Technology
  Letters \textbf{11}, 857--859 (1999).

\bibitem{zhu2016ultrabroadband}
T.~Zhu, Y.~Hu, P.~Gatkine, S.~Veilleux, J.~Bland-Hawthorn, and M.~Dagenais,
  \enquote{Ultrabroadband high coupling efficiency fiber-to-waveguide coupler
  using {Si$_3$N$_4$/SiO$_2$} waveguides on silicon,} IEEE Photonics Journal
  \textbf{8}, 1--12 (2016).

\bibitem{yao2020broadband}
Z.~Yao, Y.~Wan, Y.~Zhang, X.~Ma, and Z.~Zheng, \enquote{Broadband
  high-efficiency triple-tip spot size converter for edge coupling with
  improved polarization insensitivity,} Optics Communications \textbf{475},
  126301 (2020).

\bibitem{bhandari2020compact}
B.~Bhandari, C.-S. Im, K.-P. Lee, S.-M. Kim, M.-C. Oh, and S.-S. Lee,
  \enquote{Compact and broadband edge coupler based on multi-stage silicon
  nitride tapers,} IEEE Photonics Journal \textbf{12}, 1--11 (2020).

\bibitem{han2019polarization}
Q.~Han, J.~St-Yves, Y.~Chen, M.~M{\'e}nard, and W.~Shi,
  \enquote{Polarization-insensitive silicon nitride arrayed waveguide grating,}
  Optics letters \textbf{44}, 3976--3979 (2019).

\bibitem{hu2020ultra}
Y.~Hu, \enquote{Ultra-low-loss silicon nitride waveguide gratings and their
  applications in astrophotonics,} Ph.D. thesis, University of Maryland,
  College Park (2020).

\bibitem{zhan2023design}
J.~Zhan, Y.~Zhang, W.-L. Hsu, S.~Veilleux, and M.~Dagenais, \enquote{Design and
  implementation of a si 3 n 4 three-stigmatic-point arrayed waveguide grating
  with a resolving power over 17,000,} Optics Express \textbf{31}, 6389--6400
  (2023).

\bibitem{ishikawa2018comprehensive}
Y.~Ishikawa, M.~M. Sirk, J.~Edelstein, P.~Jelinsky, D.~Brooks, G.~Tarle,
  D.~Collaboration \emph{et~al.}, \enquote{Comprehensive measurements of the
  volume-phase holographic gratings for the dark energy spectroscopic
  instrument,} The Astrophysical Journal \textbf{869}, 24 (2018).

\bibitem{ebizuka2017novel}
N.~Ebizuka, T.~Okamoto, M.~Takeda, T.~Hosobata, Y.~Yamagata, M.~Sasaki,
  M.~Uomoto, T.~Shimatsu, S.~Sato, N.~Hashimoto \emph{et~al.}, \enquote{Novel
  gratings for next-generation instruments of astronomical observations,} in
  \enquote{Holography: Advances and Modern Trends V,} , vol. 10233 (SPIE,
  2017), vol. 10233, pp. 135--142.

\bibitem{VPH}
W.~Photonics,
  \url{https://www.spectroscopyeurope.com/system/files/pdf/WP-TN_AdvantagesVPHGratings-22Jun20-web.pdf}.
  VPH gratings by WASATCH.

\bibitem{yuan2022silicon}
S.~Yuan, J.~Feng, Z.~Yu, J.~Chen, H.~Liu, Y.~Chen, S.~Guo, F.~Huang,
  R.~Akimoto, and H.~Zeng, \enquote{Silicon nanowire-assisted high uniform
  arrayed waveguide grating,} Nanomaterials \textbf{13}, 182 (2022).

\end{thebibliography}
% bibliography data in report.bib
\bibliographystyle{osajnl} % makes bibtex use spiebib.bst

\end{document}